\documentstyle[13lomcon,cite,epsfig,psfig]{article}

\begin{document}

\title{HIGGS DECAY TO $\overline{b}b$: DIFFERENT APPROACHES TO RESUMMATION 
OF QCD EFFECTS}

\author{ A.L. Kataev \footnote{e-mail: kataev@ms2.inr.ac.ru;  
supported by RFBR Grants N 05-01-00992, 06-02-16659}} 
\address{Institute for Nuclear Research,117312  Moscow, Russia} 
\author{V.T. Kim \footnote{e-mail: kim@pnpi.spb.ru}}
\address{St. Petersburg Nuclear Physics Institute, 
188300 Gatchina,  Russia}


\maketitle\abstracts{The comparison between 
parameterisations of the perturbation results 
for the decay width of the  Standard Model Higgs boson to 
$\overline{b}b$-quarks pairs,  based on 
application of $\overline{MS}$-scheme running quark mass and pole  
 b-quark mass, are presented. In the case of the latter  parameterisation 
taking into account of  order $O(\alpha_s^3)$ term is rather important. 
It is minimising  deviations of the results obtained at the 
$O(\alpha_s^2)$ level from the results,  which follow from the  running 
quark mass approach.}

Decay widths and production cross-sections of scalar bosons are 
nowadays among the most extensively analysed theoretical quantities.
In the case if the Standard 
Electroweak Model Higgs boson has the mass   is in the region 
115 GeV$\leq  M_H\leq 2M_W$,
where the lower  bound comes from the searches of Higgs boson at
LEP2 $e^+e^-$-collider,
it can be detectable in the LHC experiments through
the mode  $H\rightarrow
\gamma\gamma$ and at Tevatron through  the main decay mode  $H
\rightarrow\overline{b}b$ (see  e.g. the  review  
\cite{Krasnikov:1999ti} ).
Moreover, the decay  $H\rightarrow \overline{b}b$ may be  seen 
at TOTEM CMS LHC experiment, which is aimed for searches of Higgs boson 
through its diffraction production (see e.g. \cite{DeRoeck:2002hk}).
The detailed study of this mode is also useful for planning
experiments at  possible future linear  $e^+e^-$-colliders \cite{Droll:2006se}.
In the mentioned region of masses 
theoretical expressions 
for  $\Gamma(H\rightarrow \overline{b}b)$ is dominating 
over expressions for other decay modes of the SM 
$H$-boson, and therefore is dominating in the denominators 
of various branching ratios, including $Br(H\rightarrow \gamma\gamma)$.
In view of all these topics  it is useful  to 
estimate theoretical error bars of $\Gamma_{H\overline{b}b}$.
To consider this question we will compare parameterisations of 
QCD predictions for $\Gamma_{H\overline{b}b}$,
 expressed through the  running $\overline{\rm MS}$-scheme     
mass $\overline{m}_b(M_H)$
 and pole quark mass $m_b$ (at the order $O(\alpha_s^2)$-effects  of perturbation theory 
the similar studies  were   made  in Ref. \cite{Kataev:1993be}).  
Consider  now the basic formula for $\Gamma_{H\overline{b}b}$  in the case 
of $N_f$=5 number of active flavours \cite{Baikov:2005rw}
\begin{eqnarray}
\Gamma_{H\overline{b}b}&=&\Gamma_0^{(b)}\frac{{\overline{m}_b}^2}{m_b^2}
\bigg(1+\Delta\Gamma_1a_s+
\Delta\Gamma_2a_s^2
+\Delta\Gamma_3a_s^3+\Delta\Gamma_4a_s^4\bigg) \\ [-5pt] 
\Delta\Gamma_1&=&17/3=5.667~,
~~\Delta\Gamma_2=d_2^{E} -\gamma_0(\beta_0+2\gamma_0)\pi^2/3 = 29.147
          \\ [-5pt] \nonumber
\Delta\Gamma_3&=&d_3^{E}-\big[d_1(\beta_0+\gamma_0)(\beta_0+2\gamma_0)
+\beta_1\gamma_0+2\gamma_1(\beta_0+2\gamma_0)\big]\pi^2/3 =41.178
\\ [-5pt] \nonumber
\Delta\Gamma_4&=&d_4^{E}-\big[d_2(\beta_0+\gamma_0)(3\beta_0+2\gamma_0)
+d_1\beta_1(5\beta_0+6\gamma_0)/2
+4d_1\gamma_1(\beta_0+\gamma_0)
\\ [-5pt]  &&{} 
+\beta_2\gamma_0+2\gamma_1(\beta_1+\gamma_1)
+\gamma_2(3\beta_0+4\gamma_0)\big]\pi^2/3\\[-5pt] \nonumber&&{}
+\gamma_0(\beta_0+\gamma_0)
(\beta_0+2\gamma_0)(3\beta_0+2\gamma_0)\pi^4/30 =-825.7 
\end{eqnarray}
where $\Gamma_0^{(b)}=\frac{3\sqrt{2}}{8\pi}G_FM_Hm_b^2$, 
$\overline{m}_b$=$\overline{m}_b(M_H)$, 
$a_s$=$a_s(M_H)$=$\alpha_s/\pi$, $\beta_i$,  $\gamma_i$ are 
the coefficients of the QCD $\beta$ and mass anomalous dimension functions 
$\Delta\Gamma_1$  was   calculated   
in \cite{Braaten:1980yq}.  
 $\Delta\Gamma_2$, $\Delta\Gamma_3$, $\Delta\Gamma_4$ 
 were  evaluated in  
\cite{Gorishnii:1990zu},\cite{Chetyrkin:1996sr},\cite{Baikov:2005rw}.
The huge  negative value  
of $\Delta\Gamma_4$  indicates, that the structure 
of perturbation series in the Minkowski region differs from the 
sign constant growth of perturbation QCD  coefficients in the Euclidean 
region. The possibility of the   manifestation of this  effect at the 
$\alpha_s^4$-level was  demonstrated previously  
 in \cite{Kataev:1995vh}, \cite{Chetyrkin:1997wm}.        
Consider the  renormalisation group (RG)  equation
\begin{equation}
\label{AD}
\overline{m}_b(M_H)=\overline{m}_b(m_b) \rm exp \bigg[-
\int_{\alpha_s(m_b)}^{\alpha_s(M_H)}
\frac{\gamma_m(x)}{\beta(x)}dx\bigg]~~~.
\end{equation}
The RG functions are known up to 4-loop level.
The solution of Eq.(\ref{AD}) is 
\begin{equation}
\overline{m}_b(M_H) = \overline{m}_b(m_b)
\bigg(\frac{a_s(M_H)}{a_s(m_b)}\bigg)^{\gamma_0/\beta_0}
\frac{AD(a_s(M_H))}{AD(a_s(m_b))}
\end{equation}
and  the coefficients of the polynomial 
$AD(a_s)$ are  expressed  through 
the coefficients of RG-functions
(see    Ref.\cite{Kataev:2001kk}). We will use  the  
QCD coupling constant 
expanded  in  inverse powers of 
${\rm ln(M_H^2/\Lambda^{(f=5)~2}_{\overline{\rm MS}}})$ 
at the NLO, NNLO and $\rm{N^3LO}$.
At the $\alpha_s^3$-level the  expression for 
$\Gamma_{H\overline{b}b}$ in terms of the quark pole mass 
 and the $\overline{\rm MS}$-scheme coupling constant can be obtained 
by three  steps.  First, one should use  the RG equation, which  translates 
$\overline{m}_b(M_H)$ to $\overline{m}_b(m_b)$. Second, 
one can use the relation  
\begin{equation}
\label{GB}
\overline{m}_b(m_b)^2=m_b^2\bigg(1-2.67a_s(m_b)-18.57a_s(m_b)^2-
175.79a_s(m_b)^3\bigg)
\end{equation}  
where the  $O(a_s^2)$-term was obtained in \cite{Gray:1990yh}
and the order $a_s^3$-term was  calculated by  semi-analytical 
methods  in \cite{Chetyrkin:1999qi} (this result was 
confirmed soon in \cite{Melnikov:2000qh} by complete analytical 
calculation.) Finally, $a_s(m_b)$ should be transformed 
to $a_s(M_H)$. 
The coefficients  
of the truncated in $a_s(M_H)$ series for $\Gamma_{H\overline{b}b}$ have  
the following numerical  forms: 
\begin{equation}
\label{OS}
\Gamma_{H\overline{b}b}=\Gamma_0^{(b)}
\bigg(1+\Delta\Gamma_1^{(OS)}a_s+
\Delta\Gamma_2^{(OS)}a_s^2
+\Delta\Gamma_3^{(OS)}a_s^3 \bigg)
\end{equation}
where 
$\Delta\Gamma_1^{(OS)}= (3-2 L)$  wih   $L\equiv ln(M_H^2/m_b^2)$ and    
\begin{equation}
\Delta\Gamma_2^{(OS)}=\bigg( -4.52 -18.139 L +0.083 L^2\bigg)
\end{equation} 
were previously  taken into account in  \cite{Kataev:1993be}, 
while we will be interested in the effect  of the next term. It reads 
\begin{equation}
\Delta\Gamma_3^{(OS)}=\bigg( -316.906 -133.421 L -1.153 L^2+0.05 L^3\bigg)~.
\end{equation}
The inclusion of the  
expressions for the two-loop diagrams 
with massive quark loop insertions, 
tabulated and taken into account in the RunDec Mathematica 
package of \cite{Chetyrkin:2000yt} leads to slight modification 
of the $a_s^2$- and $a_s^3$-corrections to Eq.(\ref{OS}): 
\footnote{Note, 
that Eq.(13) in \cite{Chetyrkin:2000yt} contains an obvious misprint 
for the 
$a_s^3$-term proportional  to $n_l^2$: instead of the first 
{\bf quadratic} term 
the {\bf linear} log term should be typed (and used).}   
 \begin{eqnarray} 
\Delta\Gamma_2^{(OS)}&=&\bigg( -5.591 -18.139 L +0.083 L^2\bigg) \\
\Delta\Gamma_3^{(OS)}&=&\bigg( -322.226 -132.351 L -1.155 L^2+0.05 L^3\bigg)
\end{eqnarray}  
The constant term of $\Delta\Gamma_3^{(OS)}$ is also affected by 
the contributions  to  the  $a_s^3$-coefficient of Eq.(6) of the  
diagrams with massive quark loop insertions, evaluated in  
\cite{Bekavac:2007tk}.  However, even in the case of charm-quark 
loop, these extra terms are  rather small. 
We will neglect these massive-dependent effects. 
Fig.1 demonstrate  the of the ratio  
$R_b(M_H)=\Gamma(H^0\rightarrow \overline{b}b)/\Gamma_0^{(b)}$
both in the case of running quark mass and pole quark mass 
parameterisations. The QCD parameters are fixed as: 
 $m_b=4.7~{\rm GeV}$ and $\overline{m}_b(\overline{m}_b)=
4.34~ {\rm GeV}$ \cite{Penin:2002zv},  
NLO: $\Lambda_{\overline{MS}}^{(5)}=253$ MeV and NNLO and 
N$^3$LO:  
 $\Lambda_{\overline{MS}}^{(5)}=220$ MeV, which correspond to the 
central values of the results,  obtained in Ref. \cite{Kataev:2001kk}

\begin{figure}[hbtp]
  \begin{center}
    \mbox{
       \epsfig{file=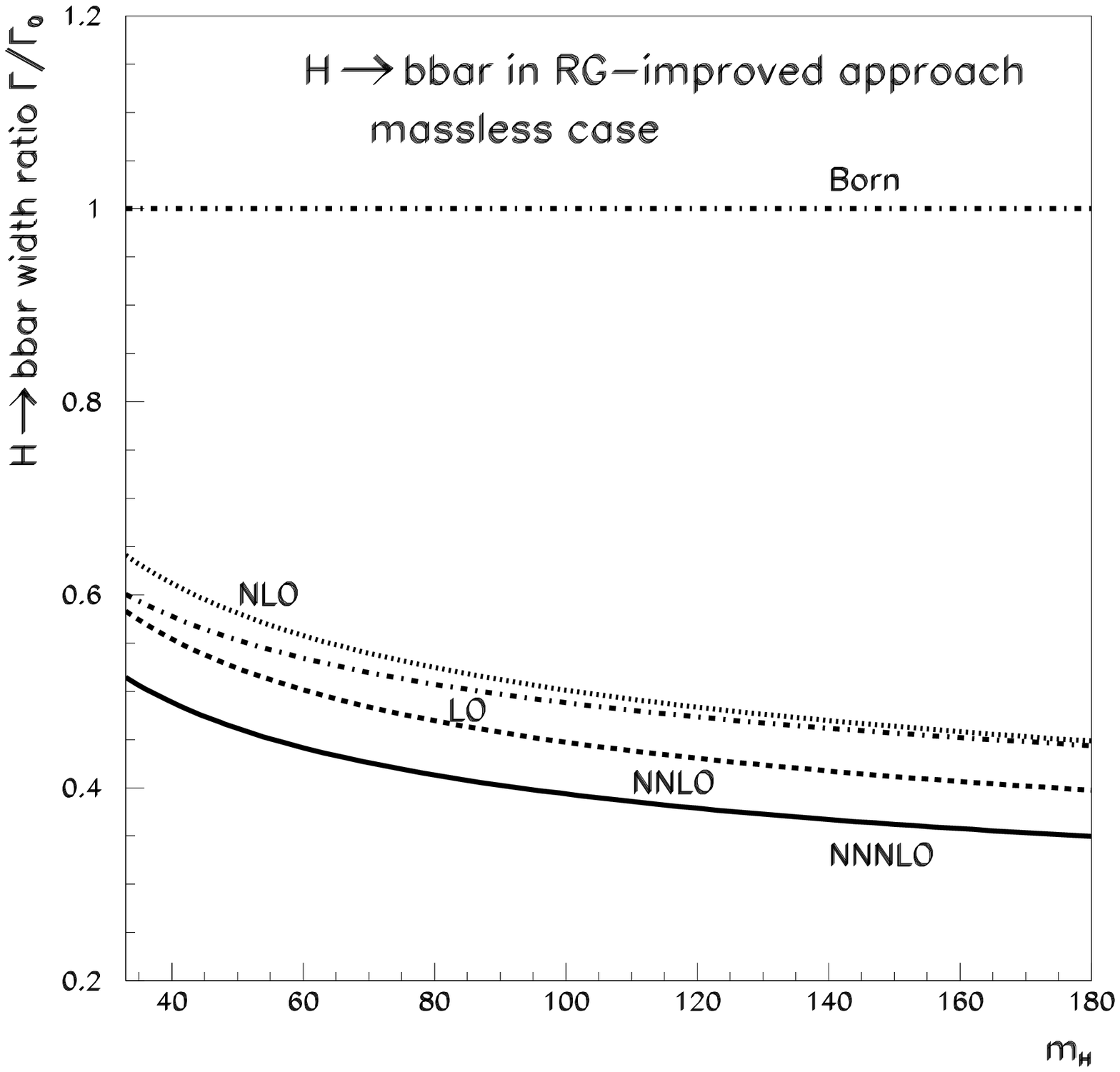,height=3.5 cm,width=5.5 cm}
       \epsfig{file=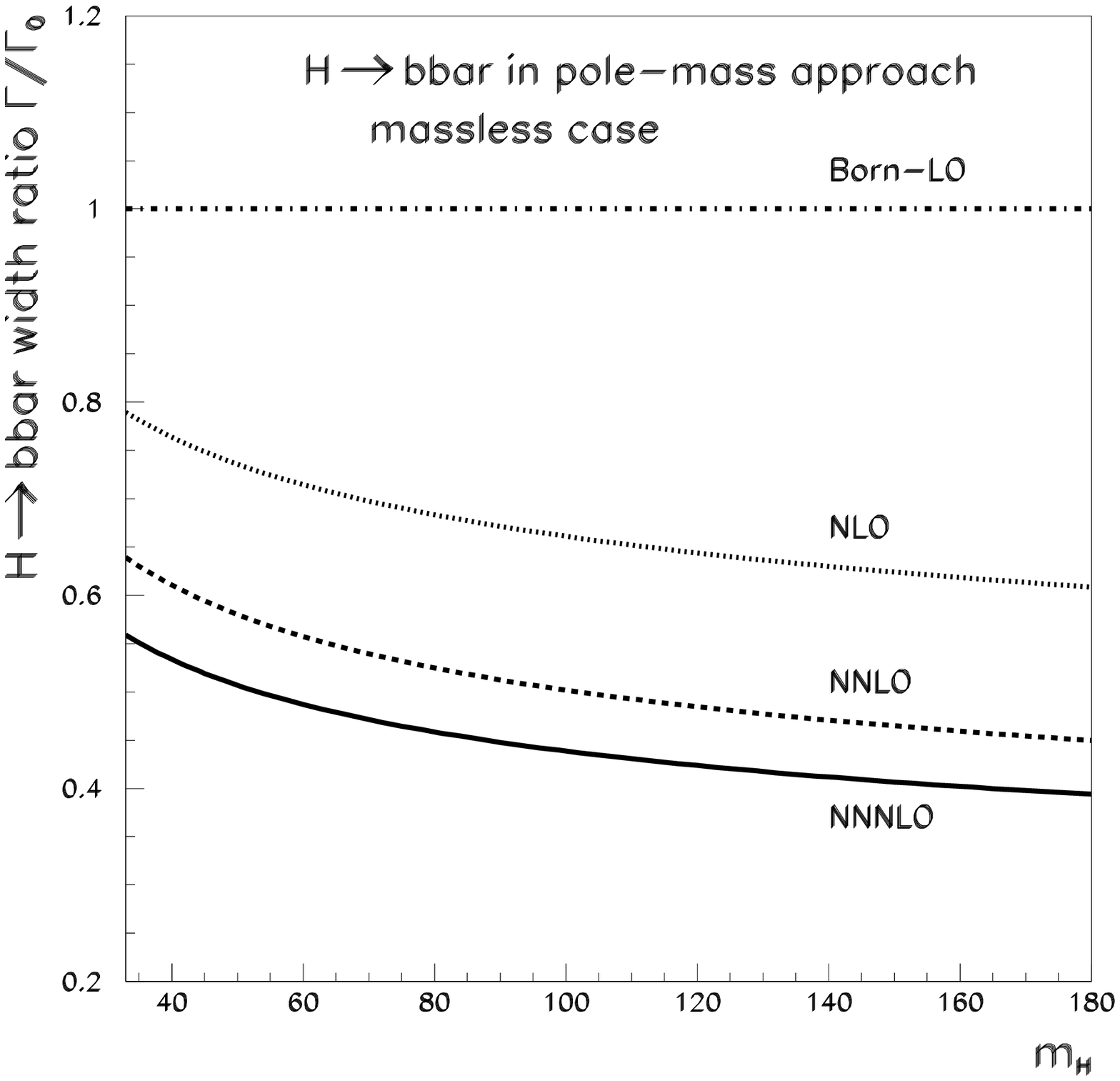,height=3.5 cm,width=5.5 cm}
     }
  \end{center}
\caption{The comparison of the expressions for  
$R_b(M_H)$. To get the impression of scale variations, 
$M_H$ is varied in the lowerer   than experimentally 
interesting region also.} 
\end{figure}

One can see, that within the application of the ``running'' quark mass
approach, or as it sometimes called RG-inspired approach, 
the results do not depend essentially from the values of the coefficient 
functions of Eq.(1), calculated within $\overline{\rm MS}$-scheme.
Almost the whole effect of decreasing of $R_b(M_H)$ ratio is described by 
effects of running of $\overline{m}_b(M_H)^{2}$. Another observation is 
that in the pole-mass parameterisation the large logs $ln(M_H^2/m_b^2)$
are important and at the $\alpha_s^3$-level their interference produce 
negative correction, which decrease the results for $R_b(M_H)$ ratio, 
making it comparable with the results of the RG-inspired 
approach.   
Another interesting problem, related to resummation of the $\pi^2$ terms 
either through fractional variant of Analytic Perturbation Theory  approach of 
Ref. \cite{Shirkov:1997wi} (see Ref. \cite{Bakulev:2006ex}) or through the 
variant of contour-improved perturbation theory of Ref. \cite{Pivovarov:1991rh}
(see Ref. \cite{Broadhurst:2000yc}),  will be considered elsewhere.

\end{document}